\documentclass[10pt,conference]{IEEEtran}
\usepackage[latin1]{inputenc}
\usepackage[cmex10]{amsmath}
\usepackage{graphicx}
\usepackage{color}
\usepackage{amsfonts,amssymb,amsthm}

\usepackage{cite}

\def\dim{\text{dim}}

\newtheorem{proposition}{Proposition}
\renewcommand{\baselinestretch}{0.98}

\IEEEoverridecommandlockouts

\title{An Algebraic Framework for Concatenated Linear Block
  Codes in Side Information Based Problems} \author{
  \IEEEauthorblockN{Felipe Cinelli Barbosa\IEEEauthorrefmark{1},
    J{\"o}rg
    Kliewer\IEEEauthorrefmark{2}, Max H.~M.~Costa\IEEEauthorrefmark{1}}\\[-1ex]
  \IEEEauthorblockA{\IEEEauthorrefmark{1}School of Electrical and
    Computer Engineering,
    University of Campinas, Campinas, Brazil\\ Email: fcb@ieee.org, max@fee.unicamp.br}
  \IEEEauthorblockA{\IEEEauthorrefmark{2}Klipsch School of Electrical
    and Computer Engineering,
    New Mexico State University, NM, USA\\ Email: jkliewer@nmsu.edu} \thanks{This work was supported in part by Funda{\c c}{\~a}o de
Amparo {\`a} Pesquisa do Estado de S{\~a}o Paulo, Brazil, under grants 2009/07239-3 and 2007/56052-8, by the Centre Interfacultaire Bernoulli, EPFL, Switzerland, and by
    the U.S.~National Science Foundation under grants CCF-0830666 and
    CCF-1017632.}  }
\IEEEaftertitletext{\vspace{-2\baselineskip}}

\begin{document}
\maketitle

\begin{abstract}

  This work provides an algebraic framework for source coding with
  decoder side information and its dual
  problem, channel coding
  with encoder side information,
  showing that nested concatenated codes can achieve the corresponding
  rate-distortion and capacity-noise bounds.
  We show that code concatenation preserves the nested properties of
  codes and that only one of the concatenated codes needs to be nested, which opens up a wide range of possible new code combinations
  for these side information based problems. In particular, the
  practically important binary version of these problems can be
  addressed by concatenating binary inner and non-binary outer linear
  codes. By observing that list decoding with folded Reed-Solomon codes is asymptotically optimal for encoding IID $q$-ary sources and that
  in concatenation with inner binary codes it can asymptotically achieve the rate-distortion bound for a
  Bernoulli symmetric source, we illustrate our findings with a new
  algebraic construction which comprises concatenated nested cyclic
  codes and binary linear block codes.


\end{abstract}

\section{Introduction}
\vspace{-1ex}
Two traditional problems in the field of communications are the
Wyner-Ziv coding problem \cite{wyner} and its dual version, the
Gel'fand-Pinsker problem \cite{gelfand}, \cite{Cos83}. The first is an instance of
distributed source coding: one source is encoded by taking advantage
of the fact that the decoder receives another correlated source as
side information. In contrast, the Gel'fand-Pinsker problem is a
channel coding problem in which a channel encoder embeds messages by
using available channel state information as side information \cite{Cos83}.
We will
refer to these problems also as source coding with decoder
side information (SCSI) for the Wyner-Ziv case, and as channel coding
with encoder side information (CCSI) for the Gel'fand-Pinsker problem
in the following, respectively.



The duality of these problems has been studied in \cite{pradhan} for
the Gaussian case, where the authors also analyze how this
relationship can be exploited to design dual coset codes.  While
nested lattice based coset coding schemes for these problems have been
proposed for continuous-input (Gaussian) channels \cite{ZSE02}, in the
following we focus on the binary version of these problems, as this is
beneficial in many applications which cope with binary data and
communication channels, as for example in digital watermarking for the case of
CCSI and in distributed video coding for SCSI.

In \cite{bennatan} superposition coding was proposed for the binary
CCSI case for which random codes and maximum-likelihood (ML) decoding
is able to achieve capacity. Differently from superposition coding,
nested codes have been used for the binary SCSI case, and a technique
based on nested parity check codes has been proposed in \cite{shamai}
which asymptotically achieves the rate-distortion bound for a
Bernoulli symmetric source.  Recently, in \cite{wainwright} the
authors present compound LDPC/LDGM constructions for both problems
which asymptotically achieve capacity for the CCSI problem and the
rate distortion bound for the SCSI problem with bounded graphical
complexity under ML decoding. They show that these compound codes
essentially have a joint source-channel coding interpretation.
Further, polar codes have been shown to be asymptotically optimal for
both problems with bounded decoding complexity \cite{KU09}. However,
their performance for practical block lengths is worse than for other
codes of the same length \cite{HUK09}. Finally, other coding schemes
for both SCSI and CCSI based on common modulation and coding schemes,
as trellis coded~quantization/modulation and turbo codes have been
presented 
(see, e.g., \cite{bennatan,liveris,Sun_etal08}).


The novel contribution of this paper is an algebraic framework
which extends the above results for the binary SCSI and CCSI cases to
concatenated nested linear block codes. In particular, we show that by
concatenating two linear block codes new binary constructions can be
obtained which preserve the nested structure either of the outer or of
the inner code. This opens up a wide range of possible new code
combinations and indicates that code concatenation can alleviate  the
search for both  practical and optimal constructions.
We analyze code concatenations for $q^m$-ary
outer codes and $q$-ary inner codes as a binary inner code can be
simply obtained by $q=2$.

Recent work by Guruswami and Rudra
\cite{guruswami} gives an explicit construction of folded RS (FRS) codes that can achieve list decoding capacity.
We show this result implies that if RS codes are used as source codes, the
rate-distortion bound is achieved for IID $q$-ary sources. Together
with the fact that concatenated binary codes using outer FRS codes can achieve list decoding capacity for concatenated codes \cite{guruswami}, it motivates the use of nested RS codes as outer codes in combination with list
decoding for both SCSI and CCSI problems.
Finally, based on our findings we exemplarily present an algebraic concatenated
nested coding scheme that asymptotically achieves the
rate-distortion and capacity-rate bounds with low encoding and
decoding complexity.

\section{Nested Linear Block Codes}
These codes were first proposed in \cite{heegard} under the name of
partitioned cyclic codes and can be generally defined as follows.

\newtheorem{definition}{Definition}
\begin{definition}[Nested Linear Block Code]
\label{def:1}
 A nested linear block code $\mathcal{C}$ is defined such that
(i) $\mathcal{C} \subset F_2^{N}$, (ii)  $\mathcal{C} = \mathcal{C}_{1} + \mathcal{C}_{2}$,
  (iii) $\mathcal{C}_{1} \cap \mathcal{C}_{2} = \{\mathbf{0}\}$,
  where $\mathcal{C}_{1}$ and $\mathcal{C}_{2}$ are subcodes.

\end{definition}

It has been shown in \cite{ZSE02, bennatan, wainwright} that nested codes are
able to achieve the rate-distortion bound for the SCSI problem and
symmetric Bernoulli sources and the capacity-noise bound for the CCSI problem
and binary symmetric channels (BSCs) as communication channels,
respectively. In the following, we revise these results and the use of
nested linear block codes in these problems,  where we focus on the binary case.

\subsection{Channel coding with encoder side information}
For this problem, we consider a BSC with  noise vector $Z \sim
\text{Bern}(p)$ (BSC($p$)) and interference $\mathbf{S}$, representing the channel
state, which is uniformly distributed over $F_2^{N}$ and  known \emph{a priori} at the encoder.
The
channel output is given by
\begin{equation}\label{GPoutput}
    \mathbf{Y} = \mathbf{E} + \mathbf{S} + \mathbf{Z},
\end{equation}
where $\mathbf{E}$ is the transmitted codeword under the input constraint
\begin{equation}\label{inputconstraint}
    \frac{1}{N} w_{H}(\mathbf{E})\leq W,
\end{equation}
with $w_H(\cdot)$ denoting the  Hamming weight.

For encoding, we assume that without loss of generality subcode
$\mathcal{C}_{1}$  carries the
information which is  transmitted in $K_1$ dimensions of the $N$-dimensional vector
space $F_2^{N}$. If $K=\dim(\mathcal{C})$ according to property
(ii) in Definition~\ref{def:1} we have that
$K_2=\dim(\mathcal{C}_{2})$ where $K = K_1 + K_2$.
Note that a nested parity-check code is simply a dual code of the
nested generator code $\mathcal{C}_G(N,K,R) =
\mathcal{C}_{1}(N-K_2,K_1,R) + \mathcal{C}_{2}(N,K_2,R+K_1)$
\cite{heegard}.

For a given information vector encoded in $\mathcal{C}_{1}$, there are
$2^{K_2}$ possible vectors in $\mathcal{C}_{2}$. The encoder now has
the task to find a vector $\mathbf{c_2}$ in $\mathcal{C}_{2}$ such that
\begin{equation}\label{S}
    \mathbf{S} = \mathbf{c_1}+ \mathbf{c_2} + \mathbf{E},
\end{equation}
 with $\mathbf{c_1} \in \mathcal{C}_{1}$, such that $\mathbf{E}$
 satisfies the constraint in (\ref{inputconstraint}). Otherwise, an
 encoder error is declared. From \eqref{GPoutput} we obtain
 the received vector as
\begin{equation}\label{Y}
    \mathbf{Y} = \mathbf{c_1}+ \mathbf{c_2} + \mathbf{Z}.
\end{equation}

\newtheorem{lemma}{Lemma}
\begin{lemma}[ \hspace{-0.6ex}\cite{wainwright}]
The  error probability in recovering $\mathbf{c_1}+ \mathbf{c_2}$
from $\mathbf{Y}$ approaches zero with increasing $N$ under the
constraint \eqref{inputconstraint} for the transmitted codeword
$\mathbf{E}$ if the maximal message rate is given as
\begin{equation}\label{GPR1}
    K_1/N = h(W)-h(p)-\epsilon.
\end{equation}
\end{lemma}

Note that  \eqref{GPR1} approaches the rates  $h(W)-h(p)$ of the capacity-noise bound
 $   R_{GP}(W,p) = \text{u.c.e.}\{h(W) - h(p),(0,0)\}$
where ``u.c.e.'' denotes the upper convex envelope. All other rates on
the curve $R_{GP}(W,p)$ can be obtained by time sharing with the point
$(0,0)$.






\subsection{Source coding with decoder side information}
This problem addresses the compression of a symmetric source $W \sim
\text{Bern}(\frac{1}{2})$ by exploiting the knowledge of another  correlated
source $Y$ as side information at the decoder. The correlation between
sources can be
represented as $W=Y\oplus S$ where $S \sim \text{Bern}(p)$ is a
``separation'' vector corresponding to errors on a virtual BSC($p$)
modeling the correlation. For the estimate of the source sequence
$\mathbf{\hat{W}}$ we require a constraint on the maximal
distortion $D$, given as
\begin{equation}\label{distortion}
    \frac{1}{N} \sum_{i=1}^{N}d_{H}(W_i,\hat{W}_i)\leq D,
\end{equation}
where $d_{H}(\cdot)$ denotes the Hamming distance.

The encoder receives a sequence of $N$ bits from source $W$,
represented by $\mathbf{W}$. It can be interpreted as
\begin{equation}\label{W}
    \mathbf{W} = \mathbf{c} + \mathbf{E},
\end{equation}
where $\mathbf{c} \in \mathcal{C}$. We also require
$\frac{1}{N}w_H(\mathbf{E})\leq D$ due to \eqref{distortion}, such that
the stored version of $W$ is given as
    $\mathbf{\hat{W}} = \mathbf{c}$,
otherwise an encoder error is declared.
We again assume that
information is  conveyed in $K_1$ dimensions of the
$N$-dimensional vector space $F_2^{N}$, corresponding to code
$\mathcal{C}_{1}$. Thus, the resulting compression rate is $K_1/N$.

At the decoder, the encoded information of length $K_1$ can be
recovered as a codeword in $\mathcal{C}_{1}$ of length $N$. Because
the decoder has access to side information $Y$ it can recover
$\mathbf{c_2}$ according to
\begin{eqnarray}\label{recc2}
  \nonumber \mathbf{c_1} + \mathbf{Y} &=& \mathbf{\hat{W}} + \mathbf{c_2} + \mathbf{W} + \mathbf{S}, \\
   &=& \mathbf{c_2} + \mathbf{E} + \mathbf{S}.
\end{eqnarray}
The decoder can then reconstruct $\mathbf{\hat{W}}$ by considering that
$\mathbf{c}=\mathbf{c_1}+\mathbf{c_2}$.

\begin{lemma}[ \hspace{-0.6ex}\cite{wainwright}]
The overall compression rate of the scheme under the distortion
constraint in \eqref{distortion} is given as
\begin{equation}
  \label{WZR1}
      K_1/N = h(p \ast D)-h(D)-\epsilon
\end{equation}
for any $\epsilon>0$, where $p \ast D = p(1-D) + D(1-p)$ represents binary convolution.
\end{lemma}
The rate $K_1/N$ in  \eqref{WZR1} approaches the rate  $h(p \ast D)-h(D)$ of
the rate-distortion bound
 $   R_{WZ}(D,p) = \text{l.c.e.}\{h(p\ast D) - h(D),(p,0)\},$
where ``l.c.e.'' denotes the lower convex envelope. All other rates on
the curve $R_{WZ}(D,p)$ can be obtained by time sharing with the point
$(p,0)$.





\section{Concatenation of Nested Codes}
The results presented in Section~II indicate that nested linear block
codes can asymptotically achieve the limits for both SCSI and CCSI
problems but does not address how practical capacity-approaching codes
for these cases can be obtained.
However, the asymptotically capacity-achieving results for compound
LDGM/LDPC codes in \cite{wainwright} suggests that code concatenation
may result in practical and efficient codes for these applications.

In this section we provide an new algebraic framework for nested concatenated
codes for which the constructions in \cite{wainwright} can be seen as
special cases. In particular, we formally prove that code concatenation
preserves the nested code structure, where the inner code serves as
translator to a $q$-ary field in such way that the outer code operates
in the corresponding $q^m$-ary extension field. This especially
also covers the practically important binary case for $q=2$.

\begin{definition} Let $\phi : F_Q^{n} \rightarrow F_q^{nm}$, with $Q=q^m$, be a bijective linear map defined as
 $   \phi(\mathbf{v}) = \mathbf{u}$,
where $\mathbf{v} \in F_Q^{n}$ and $\mathbf{u} \in F_q^{nm}$. This
means that a sequence of length $n$ in $F_Q$ can be expressed as a
$q$-ary sequence of length $nm$. If $m=1$ we have  $\mathbf{u} =
\mathbf{v}$ and there is no mapping.
\end{definition}

\begin{definition}
\label{def:3}
Let $\mathbf{u}= (\mathbf{u}_1, ..., \mathbf{u}_l)$
  , $\mathbf{u}_i \in F_q^{nm/l}, i=1,...,l$, where $1 \leq l \leq n$
  and $l$ is a divisor of $mn$. Further, let $\mathcal{C}_{\Psi}(N/l,
  nm/l, d_{\Psi})$ be a $q$-ary linear block code. Then, $\psi :
  F_q^{nm/l} \rightarrow \mathcal{C}_{\Psi}$ is a bijective linear map such that
    $\psi(\mathbf{u}_i) = \mathbf{u}_i G_{\Psi},$
where $G_{\Psi}$ is a generator matrix for $\mathcal{C}_{\Psi}$.
\end{definition}

This definition means that the sequence $\mathbf{u}$ is partitioned into
$l$ groups of $nm/l$ $q$-ary symbols that are each encoded by
$\mathcal{C}_{\Psi}$. Note that this partition
corresponds to an ($nm/l$)-folded code over $F_q^{nm/l}$. If $l=n$,
then the groups have length $m$, and are  the $q$-ary
representation of a $Q$-ary symbol.
If $l=1$, the entire $q$-ary sequence $\mathbf{u}$ is
encoded as a single input message by $\mathcal{C}_{\Psi}$.

\begin{definition}
\label{def:4}
We define the extended one-to-one linear map  $\psi^{*}:
  F_q^{nm} \rightarrow F_q^N$ as
   $ \psi^{*}(\mathbf{u}) \triangleq (\psi(\mathbf{u}_1), \dots, \psi(\mathbf{u}_l)) = (\mathbf{u}_1 G_{\Psi}, \dots, \mathbf{u}_l G_{\Psi}).$
\end{definition}

\begin{lemma}
\label{lem:3}
Let $\mathbf{v}$ be a codeword of the nested linear
  block code $\mathcal{C}(n, k, d)$ over $F_Q$ with $Q=q^m$ and
  $\mathcal{C} = \mathcal{C}_1 + \mathcal{C}_2$. The concatenation
  between $\mathcal{C}$ and $\mathcal{C}_{\Psi}(N/l, nm/l, d_{\Psi})$
  yields an equivalent code $\mathcal{C}_{eq}(N, K, D)$ over $F_q$
  according to
 $   \mathcal{C}_{eq}(N, K, D) = \{\psi^{*}(\phi(\mathbf{v}))\}_{\mathbf{v} \in \mathcal{C}}$
where $K=km$ and $D \geq d_{\Psi}d$.
\end{lemma}


The proof follows in a straightforward way from sequential concatenation \cite{mceliece}. Note that
$\phi(\mathbf{v})$ is a $q$-ary codeword of
$\mathcal{C}_q(nm, km, d_q)$, which is the $q$-ary version of
$\mathcal{C}$ in the underlying field $F_q$.  


The following proposition represents the main result of this section and
states that the nested property as stated in Definition~\ref{def:1} is
preserved if the outer code is a nested $q$-ary linear block code and
the inner code is a~$Q$-ary~linear~block~code.
\vspace{-1ex}

\begin{proposition}
\label{thm:1}
The concatenation between $\mathcal{C}$ and
  $\mathcal{C}_{\Psi}$ produces codewords of an equivalent linear code
  $\mathcal{C}_{eq}(N, K)$ over $F_q$, such that

\begin{enumerate}
  \item 
$\mathcal{C}_{eq}(N, K) = \{ \psi^{*}(\phi(\mathbf{v}_1)) \}_{\mathbf{v}_1 \in \mathcal{C}_{1}} + \{ \psi^{*}(\phi(\mathbf{v}_2)) \}_{\mathbf{v}_2 \in \mathcal{C}_{2}}$,
\vspace{0.5ex}
  \item 
$\{ \psi^{*}(\phi(\mathbf{v}_1)) \}_{\mathbf{v}_1 \in \mathcal{C}_{1}} \cap \{ \psi^{*}(\phi(\mathbf{v}_2)) \}_{\mathbf{v}_2 \in \mathcal{C}_{2}} = \{\mathbf{0}\}$.
  \end{enumerate}
\end{proposition}
\vspace{-1ex}

\begin{IEEEproof}
1) According to Lemma~\ref{lem:3} we have
$\mathcal{C}_{eq} = \{ \psi^{*}(\phi(\mathbf{v})) \}_{\mathbf{v} \in \mathcal{C}}$.
Since $\mathcal{C} = \mathcal{C}_{1} + \mathcal{C}_{2}$, then
$\mathcal{C}_{eq} = \{ \psi^{*}(\phi(\mathbf{v}_1 + \mathbf{v}_2)) \}_{\mathbf{v}_1 \in \mathcal{C}_{1}; \mathbf{v}_2 \in \mathcal{C}_{2}}$.
But as $\psi^{*}$ and $\phi$ are a linear maps and both
$\mathcal{C}_1$ and $\mathcal{C}_2$ are also subspaces over the ground
field $F_q$, as $F_Q^n$ is equivalent to $F_q^{nm}$, the additivity
property of linear mappings yields
  $\{ \psi^{*}(\phi(\mathbf{v}_1 + \mathbf{v}_2)) \}_{\mathbf{v}_1 \in
    \mathcal{C}_{1}; \mathbf{v}_2 \in \mathcal{C}_{2}} = \{
  \psi^{*}(\phi(\mathbf{v}_1)) \}_{\mathbf{v}_1 \in \mathcal{C}_{1}}
  \\+ \{ \psi^{*}(\phi(\mathbf{v}_2)) \}_{\mathbf{v}_2 \in
    \mathcal{C}_{2}}$.

\medskip
2) We define the kernel of $\psi^{*}$ to be
  $\textrm{ker} (\psi^{*}) \triangleq \{ \mathbf{u} \in \mathcal{C}_q \mid \psi(\mathbf{u}) = \mathbf{0} \}$,
where $\mathcal{C}_q(nm,km)$ is the equivalent code $\mathcal{C}$ in the
underlying field $F_q$. As $\psi^{*}$ is one-to-one, $\textrm{ker}(\psi^{*}) = \{\mathbf{0}\}$,
i.e., $\mathbf{u}=\mathbf{0}$. Equivalently, $\textrm{ker}(\phi) =
\{\mathbf{0}\}$, which means that $\mathbf{v}=\mathbf{0}$. But since
$\mathcal{C}_{1} \cap \mathcal{C}_{2} = \{\mathbf{0}\}$, then
$\mathbf{v}=\mathbf{0}$ if and only if
$\mathbf{v}_1=\mathbf{v}_2=\mathbf{0}$. In other words,
$\psi^{*}(\phi(\mathcal{C}_1 \cap \mathcal{C}_2 = \{\mathbf{0}\})) =
\{ \psi^{*}(\phi(\mathbf{v}_1)) \}_{\mathbf{v}_1 \in \mathcal{C}_{1}}
\cap \{ \psi^{*}(\phi(\mathbf{v}_2)) \}_{\mathbf{v}_2 \in
  \mathcal{C}_{2}} = \{\mathbf{0}\}$.
\end{IEEEproof}

By using a similar approach, it follows from Proposition~\ref{thm:1} that its converse also holds, i.e., that the nested property is
preserved if the inner code is a nested $q$-ary linear block code and
the outer code is a $Q$-ary linear block code.

\begin{proposition}
\label{thm:2}
The concatenation between a linear code
  $\mathcal{C}(n,k)$ over $F_Q$ and a linear code $\mathcal{C}_{\Psi}$
  over $F_q$, $\mathcal{C}_{\Psi} = \mathcal{C}_{\Psi1} +
  \mathcal{C}_{\Psi2}$ and $\mathcal{C}_{\Psi1} \cap
  \mathcal{C}_{\Psi2} = \{\mathbf{0}\}$, produces codewords of an
  equivalent linear code $\mathcal{C}_{eq}(N, K)$ over $F_q$, such
  that

\begin{enumerate}
  \item 
$\mathcal{C}_{eq}(N, K) = \{ \psi_1^{*}(\phi(\mathbf{v}))  +  \psi_2^{*}(\phi(\mathbf{v})) \}_{\mathbf{v} \in \mathcal{C}}$,
\vspace{0.5ex}
  \item 
$\{ \psi_1^{*}(\phi(\mathbf{v})) \cap \psi_2^{*}(\phi(\mathbf{v})) \}_{\mathbf{v} \in \mathcal{C}} = \{\mathbf{0}\}$,
  \end{enumerate}
where $\psi_1^{*} : F_q^{nm} \rightarrow F_q^N$ and $\psi_2^{*} : F_q^{nm} \rightarrow F_q^N$ are linear maps such that
    $\psi_j^{*}(\mathbf{u}_1,\dots, \mathbf{u}_n) = (\psi_j(\mathbf{u}_1), \dots, \psi_j(\mathbf{u}_n))$,
    with $\mathbf{u}_i \in F_q^{nm/l}$, $i=1, \dots, n$, and $\psi_j:
    F_q^{nm/l} \rightarrow \mathcal{C}_{\Psi j}$, $j = 1,2$.
\end{proposition}

Propositions~\ref{thm:1} and \ref{thm:2} show that properties (i) and
(ii) in Definition~\ref{def:1} still hold after code concatenation, no
matter whether the nested code is an inner or outer code. The only requirement is that both subcodes are concatenated with the same outer code, in order to preserve property (ii). We define the subcodes in
the resulting $q$-ary nested structure of $\mathcal{C}_{eq}$ as
$\mathcal{C}_{eq1} = \{\psi^{*}(\phi(\mathbf{v}_1))\}_{\mathbf{v}_1
  \in \mathcal{C}_1}$ and $\mathcal{C}_{eq2} =
\{\psi^{*}(\phi(\mathbf{v}_2))\}_{\mathbf{v}_2 \in \mathcal{C}_2}$
which now instead can be employed in both the SCSI and the CCSI cases.


\vspace{-1ex}
\section{Concatenated Nested Cyclic Codes and Binary Linear Block Codes}
When employing nested codes to the SCSI problem as in Section II,
$\mathcal{C}_{eq2}$ is required to be a good channel code to correct the
error formed by the source encoding distortion and the observation
error.  At the same time, $\mathcal{C}_{eq}$ must be a $D$-distortion
source code to output a codeword as close as possible to the
information sequence produced by the source with a distortion constraint
$D$. In the case of CCSI, $\mathcal{C}_{eq2}$ takes on the role of a
good $W$-distortion source code
whereas $\mathcal{C}_{eq}$ is the channel code.

While channel coding can be performed by means of good decoding
algorithms, performing source coding with error correcting codes makes it necessary to have complete algorithms that can return the nearest
codewords. Motivated by recent results on list decoding of RS codes we
will now study the suitability of these codes for source encoding.


\subsection{List decoding for folded Reed-Solomon codes}
In \cite{guruswami}, Guruswami and Rudra describe an explicit family
of codes with a list decoding algorithm that can asymptotically
achieve the information-theoretic limit of list decodability, with
encoding and decoding performed in polynomial time. The proposed codes
are folded RS codes, which can be seen as standard RS codes viewed as
codes over a larger alphabet.

\begin{definition}[$\nu$-Folded Reed-Solomon Code (FRS)] Let $\alpha
  \in F_q$ be a primitive element of $F_q$. Let $n' \leq q-1$ be a
  multiple of $\nu$ and $1 \leq k < n$. An FRS code
  $\mathcal{C}^{(\nu)}(n',k)$ over alphabet $F_q^{\nu}$ is a folded
  version of the RS code $\mathcal{C}(n,k)$ over $F_q$ and is defined
  as
\vspace{-1ex}
\begin{multline}\label{FRS}
  \{([i(\alpha^{j\nu}), i(\alpha^{j\nu+1}),...,
  i(\alpha^{j\nu+\nu-1})], 0\leq j<n') \mid \\ \textrm{deg}(i(x)) < k,
  i(x) \in F_q[x]\},
\end{multline}
\vspace{-3ex}

\noindent
where $n'=n/\nu$. In other words, a FRS code is an RS code where $\nu$
consecutive symbols each are grouped together.
\end{definition}

The GR algorithm for FRS codes of rate $R$ allows to list
decode in polynomial time up to a fraction of $(1-R-\varepsilon)$
worst-case errors. The folding operation does not change
the rate of the RS code $(R=k/n=k/n' \nu)$, thus $e'\nu/n= 1- k/n -\varepsilon$, so
$e=e'\nu$ is the number of correctable errors for the corresponding
unfolded RS code \cite{guruswami}.

\begin{proposition}\label{thm:3} If the GR list decoding algorithm is used
  in conjunction with RS codes for source encoding of IID $q$-ary sources, the probability of encoding errors asymptotically vanishes.
\end{proposition}

\begin{IEEEproof}
  Starting from an observation that the normalized covering radius
  $t(\mathcal{C})/n$ of a linear code $\mathcal{C}(n,k)$ \cite{cohen}
  is
    $t(\mathcal{C})/n \leq 1 - k/n$,
which is met with equality by RS codes, we see that
    $e = t(\mathcal{C})-\varepsilon$,
where $\varepsilon >0$, $\mathcal{C}(n,k)$ is the unfolded RS code and
$t(\mathcal{C})$ is its covering radius. Because a fraction of errors
over $F_q^{\nu}$ is equivalent to a fraction of errors over $F_q$, the
GR algorithm for FRS codes asymptotically corrects a number of errors
over $F_q$ that is equal to the covering radius of the corresponding
unfolded RS code.
\end{IEEEproof}


Note that list decoding may not output a single codeword but a list of
possible codewords. This does not pose a problem since the source
encoder can always pick the one which is closest to the source
sequence in Hamming distance.

Using concatenated codes for both the CCSI and the SCSI problems,
there are two different ways of implementing the source encoding step
of finding a vector $\mathbf{c}_2 \in \mathcal{C}_{eq2}$ and
$\mathbf{c} \in \mathcal{C}_{eq}$, respectively.  The first way is to
perform separate source encoding for each of the concatenated codes.
While RS outer codes in conjunction with the GR algorithm can optimally perform
source encoding in $F_q^n$, the performance of this strategy also depends on the
inner code.  Another way is to perform source encoding over the
concatenated binary code. In fact, list decoding capacity for binary
codes can be asymptotically achieved if FRS codes are concatenated
with random binary linear block codes (BLBC) \cite{guruswami}. This
means that every Hamming sphere of radius $h^{-1}(1-R-\varepsilon)$
has polynomially many codewords. Thus, it is possible to
asymptotically achieve the rate-distortion bound for a Bernoulli
symmetric source.



In the following we provide a general
setup which universally addresses the scenario of outer algebraic RS
or BCH codes and arbitrary inner BLBCs.

\subsection{Nested cyclic codes}
The coding scheme for the outer code is based on an algebraic
construction of nested cyclic codes. These codes form an ideal in the
polynomial ring $F_q[x]/(x^n-1)$, where $F_q[x]$ is the set of
polynomials in $x$ with coefficients from the finite field $F_q$,
where $q=2^m$. The polynomial $x^{n}-1$ can be factorized as
\begin{equation}\label{factor}
x^{n}-1 = g(x)^{(r)}f(x)^{(k_{1})}h(x)^{(k_{2})},
\end{equation}
where $r=n-(k_{1}+k_{2})$. The notation ``$a(x)^{(\cdot)}$'' is used
to indicate the degree of polynomial ``$a(x)$'', and henceforth the
argument ``$(x)$'' will be omitted in order to simplify notation. Note
that for $m=1$ we obtain BCH codes, otherwise RS codes are employed.

The polynomial $g^{(r)}$ corresponds to a generator polynomial of the code
$\mathcal{C}(n,k_{1}+k_{2})$, and $(gf)^{(r+k_{1})}$ is a generator
polynomial for $\mathcal{C}_{2}(n,k_{2})$. The codewords in $\mathcal{C}$
can be expressed as a sum of codewords as follows:
\begin{equation}\label{v}
    v^{(n-1)} = i_{1}^{(k_{1}-1)}g^{(r)} + i_2^{(k_2-1)}(gf)^{(r+k_1)},
\end{equation}
where $i_{1}^{(k_{1}-1)}g^{(r)} \in \mathcal{C}_{1}(n-k_{2},k_{1})$.

\subsection{Construction for the CCSI case}

At the outer encoder, information $i_1^{(k_1-1)}$ is encoded using
$g^{(r)}$ of code $\mathcal{C}(n,k_{1}+k_{2})$, producing
$v_1^{(n-1)}$ (zero padded to achieve length $n$) of the shortened cyclic code
$\mathcal{C}_{1}(n-k_{2},k_{1})$. In order to allow incorporating binary side
information,  the sequence $\mathbf{v}_1 \in \mathcal{C}_1$ is mapped to its
binary representation $\mathbf{u}_1$, which is then partitioned into
$l$ groups of $nm/l$ bits that are each encoded by a BLBC code
$\mathcal{C}_{\Psi}(N/l, nm/l)$. Thus, the resulting codewords $\mathbf{c}_1 =
\psi^{*}(\phi(\mathbf{v}_1))$ have length $N$ and are codewords in
$\mathcal{C}_{eq1}(N-K_2,K_1)$.

\textbf{Encoding steps:} I) (\textit{Outer encoding}): Encode
information $i_{1}^{(k_{1}-1)}$ using the generator $g^{(r)}$ for
$\mathcal{C}(n,k_1+k_2)$, thus producing a codeword $v_1^{(n-k_2-1)}$ of
$\mathcal{C}_1(n-k_2,k_1)$ padded with $k_2$ zeros; II) (\textit{Code
  concatenation}): Encode $l$ groups of $nm/l$ bits of codeword
$\mathbf{v}_1$ (received from the outer encoder) by using the inner code
$\mathcal{C}_{\Psi}$, resulting in $\mathbf{c}_1$; III) Compute
$\mathbf{S} - \mathbf{c}_1$; IV) Find $\mathbf{c}_2 \in
\mathcal{C}_{eq2}$ according to (\ref{S}) such that
(\ref{inputconstraint}) holds; V) Transmit~$\mathbf{E}$.

Note that the encoding complexity is essentially given by step IV,
because all other operations are linear. For  FRS
codes a folding/unfolding step has to be performed before finding
$\mathbf{v}_2 \in \mathcal{C}_2(n,k_2)$ as follows.

\textbf{Folding/unfolding step:} \emph{(i) Code folding $\sigma:
  F_2^{nm} \rightarrow F_{2^{m\nu}}^{n/\nu}$, $\sigma(\mathbf{u}) =
  u'$, (ii) Code unfolding \mbox{$\sigma^{*}: F_{2^{m\nu}}^{n/\nu}
  \!\rightarrow\! F_{2^{m}}^{n}, \sigma^{*}(v'_2)\!=\!v_2$.}}

\begin{proposition}\label{thm:4}
Consider a symmetric Bernoulli source. Source encoding via list
decoding of RS/BLBC code $\mathcal{C}_{eq2}$ can asymptotically achieve a vanishing
probability of encoding error. Thus, given a concatenated RS/BLBC
channel code $\mathcal{C}_{eq}$ which asymptotically achieves capacity on
the BSC($p$), the resulting joint
source-channel coding scheme for the CCSI case achieves the
capacity-noise bound $R_{GP}(W,p)$.
\end{proposition}
\begin{IEEEproof}
  From the rate distortion bound for a symmetric Bernoulli sequence,
  the rate for the $W$-distortion source code $\mathcal{C}_{eq2}$ is
  given as $K_2/N \geq 1-h(W)$.
  Because list decoding can asymptotically correct an error fraction
  of $h^{-1}(1-R_2-\varepsilon)$, we see that $R_2 =
  1-h(W)-\varepsilon$ asymptotically achieves the rate-distortion
  bound and therefore results in an encoding error probability which
  asymptotically tends to zero.  Therefore, if the RS/BLBC code
  $\mathcal{C}_{eq}$ achieves capacity on the BSC($p$), we have $R =
  1-h(p)-\varepsilon$ which results in $R_1 =
  h(W) - h(p)$. This is equivalent to the capacity-noise bound $R_{GP}(W,p)$.
\end{IEEEproof}

    The channel coding performance of the proposed scheme is
  essentially the one for the chosen concatenated RS/BLBC code
  $\mathcal{C}_{eq2}$. Here we can exploit the fact that some
  constructions (e.g., RS/LDPC) are capacity
  approaching, for which effective decoding algorithms exist.

  After transmission of $\mathbf{E}$, the decoder receives
  $\mathbf{Y}$ in (\ref{GPoutput}) and the error vector $\mathbf{Z}$
  is corrected in the same fashion as in any standard concatenated
  scheme (by using the corresponding decoding algorithms for each
  code), resulting in an error-free codeword (\ref{v}). Then, the
  embedded information $i_1$ is extracted by a modulo operation and a
  polynomial division according to
\begin{equation}\label{i1}
    i_{1}^{(k_{1}-1)} = \frac{v^{(n-1)}\mod(gf)^{(k_{1}+r)}}{g^{(r)}}.
\end{equation}

\textbf{Decoding steps:}
  I) Receive $\mathbf{Y}$, recover $\mathbf{v} \in \mathcal{C}$; II) Compute
  $i_{1}^{(k_{1}-1)}$ as in (\ref{i1}).

\subsection{Construction for the SCSI case}
The encoder receives a sequence of $N$ bits from a Bernoulli symmetric
source $W$,
represented by $\mathbf{W}$, which is equivalent to a codeword
$\mathbf{\hat{W}} = \mathbf{c}$ in $\mathcal{C}_{eq}(N,K)$ plus a
``quantization'' error $\mathbf{E}$ (\ref{W}). An encoder
error is declared if a codeword $\mathbf{\hat{W}} = \mathbf{c}$ cannot
be found.

\textbf{Encoding steps:}
  I) Receive $\mathbf{W}$, recover $\mathbf{v} \in \mathcal{C}$; II) Compute:
  $i_{1}^{(k_{1}-1)}$ as in (\ref{i1}).

  Analogous to the CCSI case, for FRS codes an extra folding/unfolding
  step must be performed before finding $\mathbf{v} \in
  \mathcal{C}(n,k)$, with the difference that now the folded codeword
  is $v' \in \mathcal{C}^{(\nu)}(n',k)$, so $\sigma^{*}(v')= v$.
  We have the following statement which is analogous to Proposition~\ref{thm:4}.

\begin{proposition}\label{thm:5}
Consider a symmetric Bernoulli source. Source encoding via list
decoding of RS/BLBC code $\mathcal{C}_{eq}$ can asymptotically achieve a vanishing
probability of encoding error. Thus, given a concatenated RS/BLBC
channel code $\mathcal{C}_{eq2}$ which asymptotically achieves capacity on
the BSC($p\ast D$), the resulting joint
source-channel coding scheme for the SCSI case achieves the
rate-distortion bound $R_{GP}(W,p)$.
\end{proposition}

The encoder extracts a polynomial $i_{1}^{(k_{1}-1)}$ of length $k_1$
($K_1$ bits) from $v^{(n-1)}$ (\ref{v}), so the compression rate is
$K_1/N$.  The encoding steps in this case are essentially the same as
the decoding steps of CCSI, but instead of channel decoding
we employ source encoding algorithms which dominate
the encoding complexity (see encoder step I).

For decoding, the steps are analogous to the encoding steps of the
CCSI case, with the difference that $\mathbf{S}$ becomes $\mathbf{Y}$,
and instead of sending the error pattern after finding $\mathbf{c}_2
\in \mathcal{C}_{eq2}$, the information word corresponding to the
actual codeword $\mathbf{\hat{W}} = \mathbf{c}_1+\mathbf{c}_2$ is
stored. Here, channel decoding is  employed which
dominates the complexity as all other
operations are linear.

\section{Conclusion}
Within the proposed algebraic framework we proved that code
concatenation preserves the nested structure of joint source-channel codes. Therefore, the
optimal asymptotic performance for both  binary SCSI and CCSI problems can be
universally achieved
by concatenation with a linear block code, provided that one of
the constituent codes has the necessary nested property.
In particular, while in \cite{wainwright} ML decoding is assumed,
through a novel RS/BLBC construction with low encoding and decoding
complexity we show that list decoding provides the optimal source
encoding performance asymptotically for both problems. At the same
time, for channel error correction any capacity-approaching algorithm
can be independently used.

It is still a challenge to exploit the full potential of concatenation
with practical list decoding algorithms, but separate source and
channel encoding is a feasible approach as  practical
encoding and decoding algorithms exist
for each code.
Future work will focus on studying other concatenated schemes
employing QC-LDPC, polar, and BCH codes as outer codes, which seems to
be a promising avenue since these codes have been
successfully employed for source coding
\cite{mceliece,matsunaga,KU09}.


\bibliographystyle{unsrt}
\bibliography{IEEEabrv,Ref2}

\end{document}